\documentclass[12pt]{iopart}

\usepackage{graphicx}
\usepackage{color}
\usepackage{multirow}

\begin{document}

\title[Iron isotope effect on $T_c$  and the crystal structure of FeSe$_{1-x}$]{Iron isotope effect on
the superconducting transition temperature and the crystal
structure of FeSe$_{1-x}$}
\author{R~Khasanov$^{1}$, M~Bendele$^{1,2}$, K~Conder$^3$, H~Keller$^2$,  E~Pomjakushina$^2$ and V~Pomjakushin$^4$}
\address{$^1$Laboratory for Muon Spin Spectroscopy, Paul Scherrer
Institute, CH-5232 Villigen PSI, Switzerland}
\address{$^2$Physik-Institut der Universit\"{a}t Z\"{u}rich,
Winterthurerstrasse 190, CH-8057 Z\"urich, Switzerland}
\address{$^3$Laboratory for Developments and Methods, Paul Scherrer Institute,
CH-5232 Villigen PSI, Switzerland}
\address{$^4$Laboratory for Neutron Scattering, ETH Z\"urich and PSI, CH-5232
Villigen PSI, Switzerland}
\ead{rustem.khasanov@psi.ch}

\begin{abstract}
The Fe isotope effect (Fe-IE) on the transition temperature $T_c$
and the crystal structure was studied in the Fe chalcogenide
superconductor FeSe$_{1-x}$ by means of magnetization and neutron
powder diffraction (NPD). The substitution of natural Fe
(containing $\simeq92$\% of $^{56}$Fe) by its lighter $^{54}$Fe
isotope leads to a shift of $T_c$ of 0.22(5)~K corresponding to an
Fe-IE exponent of $\alpha_{\rm Fe}=0.81(15)$. Simultaneously, a small structural change
with isotope substitution is observed by NDP which may contribute to the
total Fe isotope shift of $T_{c}$.

\end{abstract}


\maketitle

%
Historically, the isotope effect played  a crucial role in
elucidating the origin of the pairing interaction leading to the
occurrence of superconductivity.
The discovery of the isotope effect on the superconducting
transition temperature $T_c$ in Hg \cite{Maxwell-Reynolds50} in
1950 provided the key experimental evidence for phonon-mediated
pairing as formulated theoretically by BCS subsequently.
The observation of unusually high $T_c$'s in the newly discovered
Fe-based superconductors immediately raised the question regarding
the pairing glue and initiated isotope effect studies. Currently,
we are aware of two papers on isotope experiments with, however,
contradicting results. Liu {\it et al.} \cite{Liu09} showed that
in SmFeAsO$_{0.85}$F$_{0.15}$ and Ba$_{0.6}$K$_{0.4}$Fe$_2$As$_2$
the Fe isotope effect (Fe-IE) exponent,
\begin{equation}
\alpha_{\rm Fe}=-{\rm d}\ln T_c/{\rm d}\ln M_{\rm Fe} = - (\Delta T_{c})/T_{c})/(\Delta M_{\rm Fe}/M_{\rm Fe}),
 \label{eq:alpha}
\end{equation}
reaches values of $\alpha_{\rm Fe}\simeq0.35$ ($M_{\rm Fe}$ is the
Fe atomic mass), while Shirage {\it et al.} \cite{Shirage09} found
a negative Fe-IE exponent $\alpha_{\rm Fe}\simeq -0.18$ in
Ba$_{1-x}$K$_{x}$Fe$_2$As$_2$. Note, that the only difference
between the Ba$_{1-x}$K$_{x}$Fe$_2$As$_2$ samples studied in
Refs.~\cite{Liu09} and \cite{Shirage09} was the preparation
procedure (low-pressure synthesis in \cite{Liu09} {\it vs.}
high-pressure synthesis in \cite{Shirage09}), while the potassium
doping ($x\simeq0.4$) as well as the $T_c$'s for the samples
containing natural Fe ($T_c\simeq 37.3$~K in \cite{Liu09} {\it
vs.} $T_c\simeq37.8$~K in \cite{Shirage09}) were almost the same.

In this paper we study the Fe-IE on $T_c$ and on
the structural parameters (such as the lattice parameters $a$,
$b$, and $c$, the lattice volume $V$, and the distance between the
Se atom and Fe plane, Se height $h_{\rm Se}$) for another
representative of the Fe-based high-temperature superconductors
(HTS), namely FeSe$_{1-x}$. The substitution of
natural Fe (containing $\simeq92$\% of $^{56}$Fe) by its lighter
$^{54}$Fe isotope leads to a shift of $T_c$ of 0.22(5)~K
corresponding to an Fe-IE exponent of $\alpha_{\rm Fe}=0.81(15)$.

%
The $^{54}$FeSe$_{1-x}$/$^{56}$FeSe$_{1-x}$ samples (here after we denote
natural Fe containing $\simeq92$\% of $^{56}$Fe isotope as $^{56}$Fe) with the
nominal composition FeSe$_{0.98}$ were prepared by a solid state reaction made
in two steps.
Pieces of Fe (natural Fe: 99.97\% minimum purity, average atomic mass
$M_{Fe}=55.85$~g/mol, or $^{54}$Fe: 99.99\% purity, 99.84\% isotope enriched,
$M_{^{54}Fe}=54.0$~g/mol) and Se (99.999\% purity) were first sealed in double
walled quartz ampules, heated up to 1075$^{\rm o}$C, annealed for 72~h at this
temperature and 48~h at 420$^{\rm o}$C, and then cooled down to room
temperature at a rate of 100$^{\rm o}$C/h. As a next step, the samples, taken
out of the ampules, were powderised, pressed into pellets, sealed into new
ampules and annealed first at 700$^{\rm o}$C for 48~h and then at 400$^{\rm
o}$C for 36~h, followed by cooling to room temperature at a rate of 200$^{\rm
o}$C/h. Due to the extreme sensitivity of FeSe$_{1-x}$ to oxygen
\cite{Pomjakushina09}, all the intermediate steps (grinding and pelletizing) as
well as the preparation of the samples for the neutron powder diffraction and
magnetization experiments were performed in a glove box under He atmosphere.

%

The Fe-IE on the structural properties was studied by neutron powder
diffraction (NPD) experiments by using the high-resolution powder
diffractometer HRPT (Paul Scherrer Institute, Switzerland)
\cite{Fischer_HRPT_00}. The experiments were carried out at a wavelength
$\lambda = 1.494$~{\AA}. The $^{54}$FeSe$_{1-x}$/${^{56}}$FeSe$_{1-x}$ samples,
placed into vanadium containers, were mounted into a He-4 cryostat in order to
reach temperatures between 5 and 250~K. High statistics data were taken at 250
and 5~K. Data at $10\leq T\leq240$~K were collected with intermediate
statistics.

%
\begin{figure}[htb]
 \begin{center}
\includegraphics[width=0.7\linewidth]{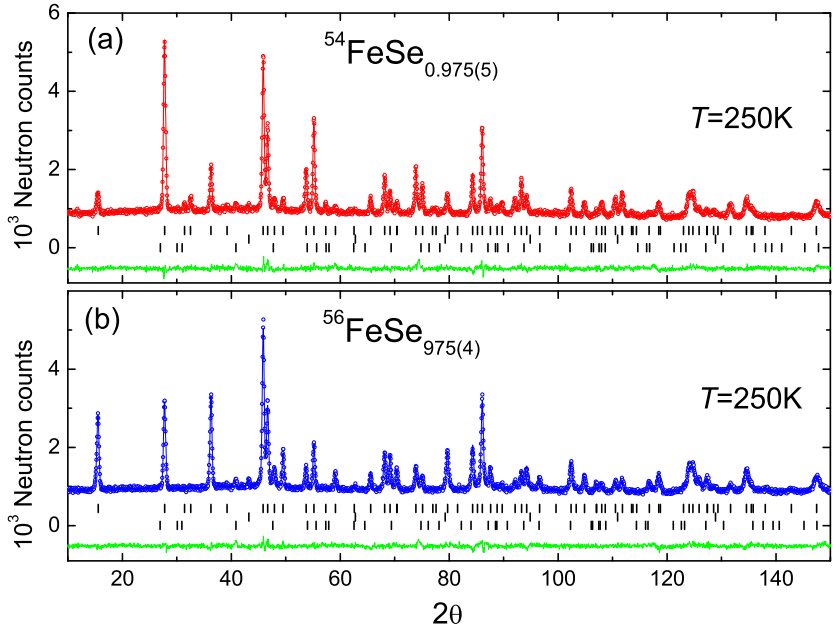}
 \end{center}
%
\caption{(Color online) The Rietveld refinement pattern and difference plot of
NPD data for $^{54}$FeSe$_{1-x}$ (panel a) $^{56}$FeSe$_{1-x}$ (panel b) at
$T=250$~K. The rows of ticks show the Bragg-peak positions for the main phase
FeSe ($P4nmm$) and two impurity phases: Fe ($Im3m$) and hexagonal FeSe
($P6_3/mmc$). The main tetragonal phase corresponds to 0.975(5) and 0.975(4) Se
occupancy for $^{54}$FeSe$_{1-x}$ and $^{56}$FeSe$_{1-x}$, respectively. }
 \label{fig:NPD-spectra}
\end{figure}
%

Figure~\ref{fig:NPD-spectra} shows the NPD spectra taken at
$T=250$~K. The differences in peak intensities, clearly visible at
small $\theta$, are caused by the different values of the coherent
neutron scattering length ($b_{coh}$) of natural Fe and that of
the $^{54}$Fe isotope. The refinement of the crystal structure was
performed by using the FULLPROF program \cite{FULLPROF} with
$b_{coh}^{Fe}=9.45\cdot 10^{-15}$~m, $b_{coh}^{\
{^{54}Fe}}=4.2\cdot 10^{-15}$~m, and $b_{coh}^{Se}=7.97\cdot
10^{-15}$~m \cite{nist}. The refined structural parameters at $T=250$~K and 5~K are
summarized in Table~1. The amount of the impurity phases and the Se content ($1-x$),
determined for the data sets taken at $T=250$~K, were kept fixed
during the refinement of the NPD spectra at lower temperatures.
The mass fractions of impurity phases, the hexagonal FeSe ($P6_3/mmc$) and Fe
($Im3m$), were found to be 0.50(10)\%, 0.31(4)\% and 1.13(18)\%, 1.06(7)\% for
$^{54}$FeSe$_{1-x}$ and $^{56}$FeSe$_{1-x}$, respectively.

%
\begin{table}[htb]
\caption[~]{\label{Table:NPD_IE-results} Structural parameters of
$^{54}$FeSe$_{1-x}$ and $^{56}$FeSe$_{1-x}$ at $T=250$ and 5~K. Space group
$P4/nmm$ (no.~129), origin choice 2: Fe in ($2b$) position (1/4, 3/4, 1/2); Se
in ($2c$) position (1/4, 1/4, $z$). Space group $Cmma$ (no.~67): Fe in ($4b$)
position (1/4, 0, 1/2), Se in ($4g$) position (0, 3/4, $z$). The atomic
displacement parameters ($B$) for Fe and Se were constrained to be the same.
The Bragg $R$ factor is given for the main phase; the other reliability factors
are given for the whole refinement.
}
\begin{center}
\begin{tabular}{lcc|cc}\\
 \hline
 \hline
& \multicolumn{2}{c}{$T=250$~K}& \multicolumn{2}{c}{$T=5$~K}\\
&$^{54}$FeSe$_{1-x}$ & $^{56}$FeSe$_{1-x}$ & $^{54}$FeSe$_{1-x}$ & $^{56}$FeSe$_{1-x}$\\
%
%
\hline
Space group &  \multicolumn{2}{c|}{$P4/nmm$}& \multicolumn{2}{c}{$Cmma$}\\
Se content& 0.975(5) & 0.975(4)&\multicolumn{2}{c}{fixed to 0.975}\\
$a$({\AA})&\multirow{2}{*}{3.77036(3)}&\multirow{2}{*}{3.76988(5)}
&5.33523(10)&5.33426(10)\\
$b$({\AA})&&&5.30984(10)&5.30933(10)\\
$c$({\AA})&5.51619(9)&5.51637(9)&5.48683(9)&5.48787(9)\\
Volume (\AA$^3$)&156.883(3)&156.797(3)&155.438(5)&155.424(5)\\
$z-$Se&0.2319(2)&0.2326(0.3)&0.2321(2)&0.2322(3)\\
$B$({\AA}$^2$)&1.02(2)&0.93(2)&0.44(2)&0.36(2)\\
$R_{\rm Bragg}$&3.11&2.93&4.13&3.63\\
$R_{wp}$&3.93&3.72&5.16&4.62\\
$R_{exp}$&3.13&3.05&4.73&4.03\\
$\chi^2$&1.58&1.49&1.19&1.32\\
 \hline \hline \\

\end{tabular}
   \end{center}
\end{table}
%


%
\begin{figure}[htb]
 \begin{center}
\includegraphics[width=0.7\linewidth]{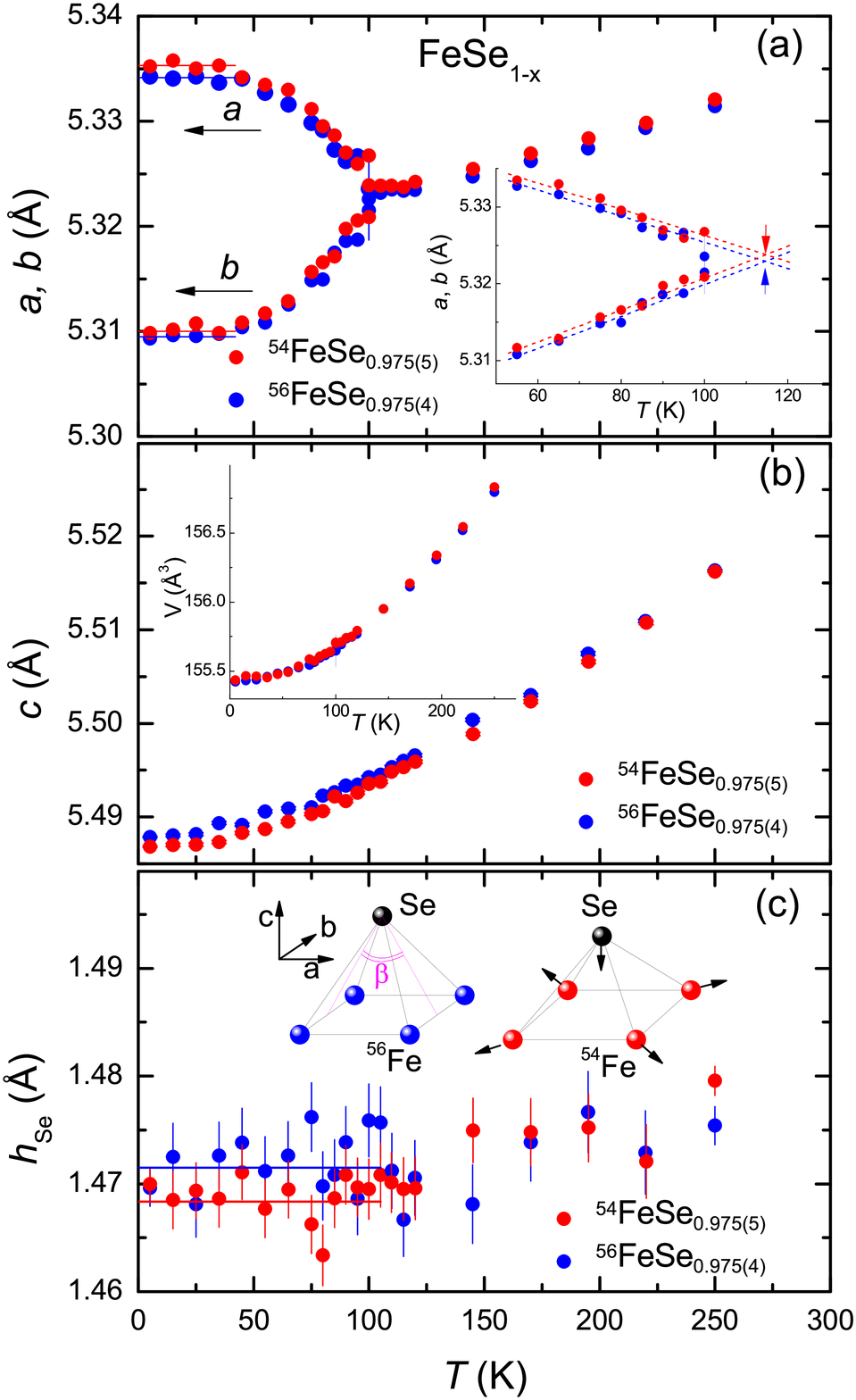}
 \end{center}
%
\caption{(Color online) The temperature dependence of the lattice constants $a$
and $b$ (panel a, in the tetragonal phase $a$ is multiplied by $\sqrt{2}$),
lattice constant $c$ (panel b), and the distance between the Se atom and Fe
plane $h_{\rm Se}$ (panel c) for $^{54}$FeSe$_{0.975(5)}$ and
$^{56}$FeSe$_{0.975(4)}$ samples. The inset in panel a shows the extended part
of $a(T)$ and $b(T)$ in the vicinity of the structural transition temperature
$T_s$, together with the linear fits. The inset in panel b represents the
temperature dependence of the lattice volume $V$. The inset in panel c shows
schematically the modification of the Fe$_4$Se pyramid caused by
$^{56}$Fe/$^{54}$Fe isotope substitution. The arrows show the direction of atom
displacements (see text for details).}
 \label{fig:NPD-lattice_parameters}
\end{figure}
%

\begin{figure}[htb]
 \begin{center}
\includegraphics[width=0.7\linewidth]{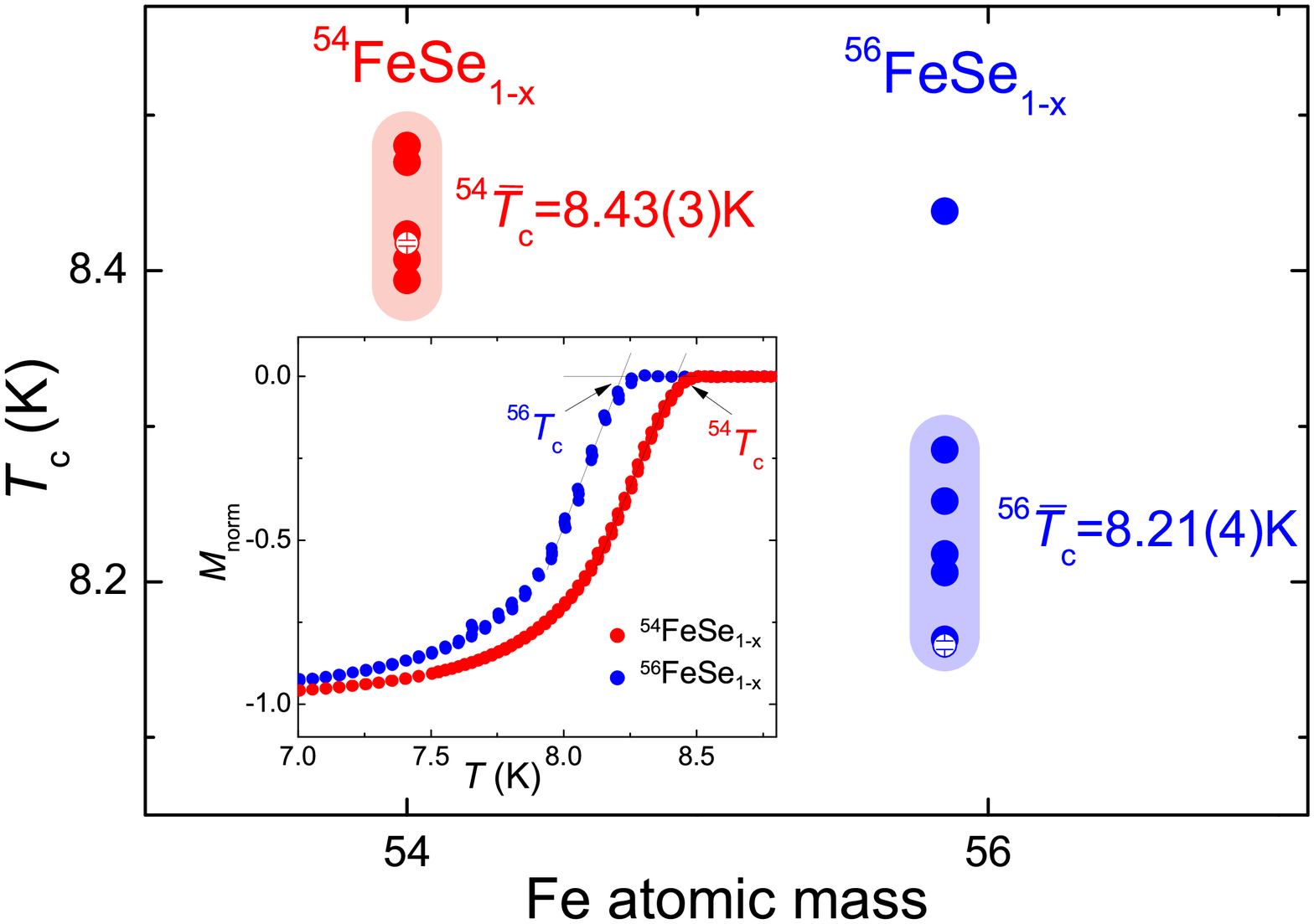}
 \end{center}
%
\caption{(Color online) The superconducting transition temperature $T_c$ as a
function of Fe atomic mass for $^{54}$FeSe$_{1-x}$/$^{56}$FeSe$_{1-x}$ samples
studied in the present work. The open symbols correspond to the samples studied
by NPD experiments. The $T_c$'s fall into the regions marked by the colored
stripes with the corresponding mean values $^{54}\overline{T}_c=8.43(3)$~K and
$^{56}\overline{T}_c=8.21(4)$~K. The inset shows the normalized ZFC
magnetization curves $M_{norm}(T)$ for one pair of $^{54}$FeSe$_{1-x}$ and
$^{56}$FeSe$_{1-x}$ samples. The transition temperature $T_c$ is determined as
the intersection of the linearly extrapolated $M_{nomr}(T)$ curve in the
vicinity of $T_c$ with the $M=0$ line.}
 \label{fig:ZFC-DC}
\end{figure}

%
Figure~\ref{fig:NPD-lattice_parameters} shows the temperature dependence of the
lattice parameters $a$, $b$, and $c$, the lattice volume $V$, and the Se height
$h_{\rm Se}$ of a representative $^{54}$FeSe$_{1-x}$ and a representative
$^{56}$FeSe$_{1-x}$ sample (see Fig.~\ref{fig:ZFC-DC}). From
Fig.~\ref{fig:NPD-lattice_parameters}a it is obvious that at $T_{s}\simeq100$~K
a transition from a tetragonal to an orthorhombic structure takes place,
analogous  to that reported  in \cite{Pomjakushina09,Margadonna08}. The Fe-IE
on the structural transition temperature $T_s$ could be estimated from the
shift of the interception point of the linear fits to $a(T)$ and $b(T)$ in the
vicinity of $T_s$, as denoted by the arrows in the inset of
Fig.~\ref{fig:NPD-lattice_parameters}a, which was found to be $\Delta
T_s=0.2(2.5)$~K. Within the whole temperature range (5~K$\leq T\leq$250~K) the
lattice constants $a$ and $b$ are slightly larger for $^{54}$FeSe$_{1-x}$ than
those for $^{56}$FeSe$_{1-x}$ (see Fig.~\ref{fig:NPD-lattice_parameters}a).
This is in contrast to the lattice parameter $c$, which within the same range
is marginally smaller for $^{54}$FeSe$_{1-x}$ than for $^{56}$FeSe$_{1-x}$
(Fig.~\ref{fig:NPD-lattice_parameters}b). The lattice volume remains, however,
unchanged. Consequently, substitution of $^{56}$Fe by $^{54}$Fe leads to a
small, but detectable {\it enhancement} of the lattice along the
crystallographic $a$ and $b$ directions and a {\it compression} of it along the
$c-$axis, resulting in a change of the shape of the Fe$_4$Se pyramid, which is
known to influence $T_c$ in Fe-based HTS \cite{Zhao08,Horigane09,Mizuguchi10}.
This is shown in Fig.~\ref{fig:NPD-lattice_parameters}c where below 100~K the
Se atom is located closer to the Fe plane in $^{54}$FeSe$_{1-x}$ than in
$^{56}$FeSe$_{1-x}$. The corresponding change of the Fe$_4$Se pyramid is shown
schematically in the inset of Fig.~\ref{fig:NPD-lattice_parameters}c. It is
important to note that the observed Fe-IE's on the lattice parameters are
intrinsic and not just a consequence of slightly different samples. As shown in
Ref.~\cite{Pomjakushina09}, various samples of $^{56}$FeSe$_{1-x}$  with $1-x
\simeq 0.98$ and $T_{c} \simeq 8.2$~K indeed exhibit the same lattice
parameters within experimental error.



%
%


The Fe-IE on the transition temperature $T_c$ was studied by means of magnetization experiments.
Measurements were performed by using a SQUID magnetometer (Quantum Design
MPMS-7) in a field of $\mu_0H=0.1$~mT for temperatures ranging from 2 to 20~K.
In order to avoid artifacts and systematic errors in the determination of the isotope shift of
$T_{c}$ it is important to perform a {\it statistical} study: {\it i.e.} to
investigate series of $^{54}$FeSe$_{1-x}$/${^{56}}$FeSe$_{1-x}$
samples synthesized exactly the same way (the same thermal
history, the same amount of Se in the initial composition). The
magnetization experiments were conducted for six
$^{54}$FeSe$_{1-x}$ and seven $^{56}$FeSe$_{1-x}$ samples,
respectively.
The inset in Fig.~\ref{fig:ZFC-DC} shows an example of zero-field cooled (ZFC)
magnetization curves for a pair of $^{54}$FeSe$_{1-x}$/$^{56}$FeSe$_{1-x}$
samples ($M_{norm}$ was obtained after subtracting the small paramagnetic
offset $M_{magn}$ measured at $T>T_c$ and further normalization of the obtained
curve to the value at $T=2$~K, see Fig.~1 in Ref.~\cite{Pomjakushina09} for
details). The magnetization curve for $^{54}$FeSe$_{1-x}$ is shifted almost
parallel to higher temperature, implying that $T_c$ of $^{54}$FeSe$_{1-x}$ is
higher than that of $^{56}$FeSe$_{1-x}$.
The resulting transition temperatures determined from the
intercept of the linearly extrapolated $M_{norm}(T)$ curves with
the $M=0$ line for all samples investigated are summarized in
Fig.~\ref{fig:ZFC-DC}. The $T_c$'s for both sets of
$^{54}$FeSe$_{1-x}$/$^{56}$FeSe$_{1-x}$ samples fall into two
distinct regions: $8.39\leq {^{54}T_c} \leq 8.48$~K and $8.15\leq
{^{56}T_c} \leq 8.31$~K, respectively. The corresponding mean
values are: $^{54}\overline{T}_c=8.43(3)$~K and
$^{56}\overline{T}_c=8.21(4)$~K. Note, that one out of the seven
$^{56}$FeSe$_{1-x}$ samples had $T_c\simeq 8.44$~K which is by
more than 5 standard deviations above the average calculated for
the rest of the six samples. We have no explanation for this
discrepancy, but decided to show this point for completeness of
the data collected.

The Fe-IE exponent $\alpha_{\rm Fe}$ was determined from the data presented in Fig.~\ref{fig:ZFC-DC} using Eq.~(\ref{eq:alpha}),
where the relative Fe isotope shift of the quantity $X$ is defined as $\Delta X/X=(^{54}X-^{56}X)/^{56}X$ (this definition of
$\Delta X/X$ is used throughout the paper).
With $^{54}\overline{T}_c=8.43(3)$~K, $^{56}\overline{T}_c=8.21(4)$~K, $M_{^{54}\rm Fe} = 54$~g/mol, and $M_{^{56}\rm Fe} = 55.85$~g/mol one obtains $\alpha_{\rm Fe}= 0.81(15)$.
Two points should be emphasized:
i) The {\it positive} sign of the Fe-IE exponent $\alpha_{\rm Fe}$
is similar to that observed in phonon mediated superconductors,
such as elemental metals \cite{Maxwell-Reynolds50} and MgB$_2$
\cite{Budko01-Hinks01} as well as in cuprate HTS
\cite{Batlogg87-Franc90,Khasanov08_OIE-YPr} where the pairing
mechanism is still under debate. Bearing in mind that a positive
Fe-IE exponent was also observed  in SmFeAsO$_{0.85}$F$_{0.15}$
and Ba$_{0.6}$K$_{0.4}$Fe$_2$As$_2$ \cite{Liu09}, we may conclude
that at least for three compounds representing different families
of Fe-based HTS (1111, 122, and 11) the sign of the Fe-IE on $T_c$
is conventional. This suggests that the lattice plays an essential role in
the pairing mechanism in the Fe-based HTS.
ii) The Fe-IE exponent $\alpha_{\rm Fe}= 0.81(15)$ is larger than
the BCS value $\alpha^{\rm BCS}=0.5$ as well as more than twice as
large as $\alpha_{\rm Fe}\simeq0.35$ reported for
SmFeAsO$_{0.85}$F$_{0.15}$ and Ba$_{0.6}$K$_{0.4}$Fe$_2$As$_2$
\cite{Liu09}. Note that an enhanced value of the oxygen isotope
exponent ($\alpha_{\rm O}\simeq1$) was also observed in underdoped
cuprate HTS \cite{Khasanov08_OIE-YPr} and was shown to be a
consequence of the polaronic nature of the supercarriers in that
class of materials \cite{Bussmann-Holder05}. Recently,
Bussmann-Holder {\it et al.} \cite{Bussmann-Holder09-1-2}
showed that in the framework of a two-band model polaronic coupling in the larger gap
channel as well as in the interband interaction induce a $T_{c}$ (doping)
dependent Fe-IE: $\alpha_{\rm Fe}$ increases strongly with reduced $T_{c}$ (doping), reaching
$\alpha_{\rm Fe} \simeq  0.9$ at $T_{c} \simeq 10$~K. Note that a similar generic trend is observed
in cuprate HTS \cite{Batlogg87-Franc90,Khasanov08_OIE-YPr}.

However, our structural refined NPD data suggest that part of the large Fe-IE
$\alpha_{\rm Fe}= 0.81(15)$ may result from the tiny structural
changes due to $^{54}$Fe/$^{56}$Fe substitution.
In the following we discuss a possible structural effect on the observed Fe-IE on $T_{c}$.
It is known that in FeSe$_{1-x}$ a decrease of the Se height caused by
compression of the Fe$_4$Se pyramid leads to an increase of $T_c$
by $\Delta T_c^{h_{\rm Se}}/(\Delta h_{\rm Se}/h_{\rm Se})\simeq
3.4$~K/\% \cite{Mizuguchi10,Margadonna09}. In contrast, an
increase of the Se(Te)-Fe-Se(Te) angle in the FeSe$_{1-y}$Te$_{y}$
family (angle $\beta$ in our notation \cite{comment}, see the
inset of Fig.~\ref{fig:NPD-lattice_parameters}c) results for
$y\leq0.5$ in a decrease of $T_c$ by $\Delta T_c^\beta/(\Delta
\beta/\beta)\simeq 2.9$~K/\% \cite{Horigane09}. Based on the
structural data presented in Fig.~\ref{fig:NPD-lattice_parameters}
one obtains $\Delta h_{\rm Se}/h_{\rm Se}=0.22(8)$\% and
$\Delta\beta/\beta=-0.13(4)$\%, leading to $\Delta T_c^{h_{\rm
Se}}=0.7(3)$~K and $\Delta T_c^\beta=-0.4(2)$~K (in this estimate
the values of the lattice constants $a$ and $b$, and $h_{\rm Se}$
were averaged over the temperature regions denoted as solid lines
in Figs.~\ref{fig:NPD-lattice_parameters}a and c).
Bearing in mind that all Fe-based HTS are similarly sensitive to
structural changes as FeSe$_{1-x}$ (see, {\it e.g.},
\cite{Zhao08,Horigane09,Mizuguchi10,Mizuguchi09}) we conclude that the shift
of $T_c$ caused by tiny modifications of the crystal structure
upon isotope exchange may contribute to
the total Fe-IE exponent. However, the large errors of $\Delta
T_c^{h_{\rm Se}}$ and $\Delta T_c^\beta$ do not allow a reliable estimate
of this structural effect on the Fe-IE on $T_{c}$.
A better estimate of this effect can be made
based on the empirical relation between $T_{c}$ and the lattice parameter
$a$ for the 11 family FeSe$_{1-y}$Te$_{y}$ \cite{Horigane09,Mizuguchi09}. Assuming that the relation $T_{c}(a)$ is also valid for FeSe$_{1-x}$ one obtains from the data presented in Ref.~\cite{Mizuguchi09} for $y\leq 0.5$ the relation
$\Delta T_c^a/(\Delta a/a) \approx 6$~K/\%.
With $(\Delta a +\Delta b)/(a+b)=0.0195(14)$\% this gives rise to a {\em structural}
shift of $T_c$ of $\Delta T_c^{\rm str}\approx 0.1$~K
(the lattice constants $a$ and $b$ were averaged over the
temperature regions marked as solid line in Fig.~\ref{fig:NPD-lattice_parameters}a). Taking this correction into account yields a rough estimate of the intrinsic Fe-IE exponent of $\alpha_{\rm Fe}^{\rm int} \approx 0.4$. This value is comparable with
$\alpha_{\rm Fe}\simeq 0.35$ reported for SmFeAsO$_{0.85}$F$_{0.15}$ and Ba$_{0.6}$K$_{0.4}$Fe$_2$As$_2$ \cite{Liu09}.

%
%

To summarize, the $^{56}$Fe/${^{54}}$Fe isotope effects on the superconducting
transition temperature and the crystal structure were studied in the iron
chalcogenide superconductor FeSe$_{1-x}$. The following results were obtained:
(i) The substitution of the natural Fe ($M_{\rm Fe}=55.85$~g/mol) by the
$^{54}$Fe isotope ($M_{^{54}{\rm Fe}}=54.0$~g/mol) gives rise to a pronounced Fe isotope shift of the transitions temperature $T_{c}$ as determined by magnetization measurements. The average $T_{c}$ is found to be $\simeq
0.22$~K higher for the $^{54}$FeSe$_{1-x}$ samples as compared to
the $^{56}$FeSe$_{1-x}$ samples resulting in a Fe-IE exponent of
$\alpha_{\rm Fe}=0.81(15)$.
(ii) The $^{56}$Fe/$^{54}$Fe isotope substitution leads to an enhancement of
the lattice constants $a$ and $b$ and a shrinkage of the lattice constant $c$.
These modifications do not affect the lattice volume.
(iii) The tetragonal to orthorhombic structural transition
temperature ($T_s\simeq 100$~K) is the same for both
$^{54}$FeSe$_{1-x}$ and $^{56}$FeSe$_{1-x}$ within the accuracy of
the experiment.
(iv) For temperatures below 100~K the distance between the Se
atom and Fe plane (Se height) is smaller for the samples with
$^{54}$Fe. This, together with the results of point ii), imply that
$^{56}$Fe/$^{54}$Fe isotope substitution leads to a compression
of the Fe$_4$Se pyramid along the crystallographic $c-$axis and an
enhancement along the $a-$ and $b-$directions.
(v) The structural changes caused by $^{56}$Fe/$^{54}$Fe
isotope substitution induce a shift in $T_c$ which may
reduce the value of Fe-IE exponent to $\approx 0.4$, in fair agreement with
$\alpha_{\rm Fe}\simeq 0.35$ obtained for
SmFeAsO$_{0.85}$F$_{0.15}$ and Ba$_{0.6}$K$_{0.4}$Fe$_2$As$_2$
\cite{Liu09}.

In conclusion, from magnetization experiments the Fe-IE exponent of $T_{c}$ for the FeSe$_{1-x}$
system was determined to be $\alpha_{\rm Fe}=0.81(15)$. The tiny changes of the
structural parameters caused by isotope substitution may contribute to the total Fe-IE exponent,
and may help to clarify or even be the origin of the previously reported
controversial results \cite{Liu09,Shirage09}. However, more detailed
and systematic structural investigations on Fe isotope substituted samples are required
in order to draw definite conclusions. Our findings, on the other hand, clearly
show that a conventional isotope effect on $T_c$ is present which
highlights the role of the lattice in the pairing mechanism in this new material class.

We would like to thank A. Bussmann-Holder for fruitful discussions
and for the critical reading of the manuscript.
This work was partly performed at SINQ (Paul Scherrer Institute,
Switzerland). The work of MB was supported by the Swiss National
Science Foundation. The work of EP was supported by the NCCR
program MaNEP.


\begin{thebibliography}{99}
\bibitem{comment} In the orthorombic phase there are two angles $\beta_1$ and
$\beta_2$ which are different by $\simeq0.3^{\rm o}$ at $T=5$~K.
\end{thebibliography}
\end{document}